\documentclass[%
 amsmath,amssymb,
superscriptaddress,
aps,pra,
twocolumn
]{revtex4-1}

\usepackage{verbatim}
\usepackage{graphicx}
\usepackage{braket}
\usepackage{amsmath}
\usepackage{siunitx}
\usepackage{textcomp,gensymb}
\usepackage{braket}
\usepackage{upgreek}
\usepackage{graphicx}
\usepackage{dcolumn}
\usepackage{bm}
\usepackage{natbib}

\setcitestyle{super}

\begin{document}


\title{Experimental entangled photon pair generation using crystals with parallel optical axes}

\author{Aitor Villar}
\affiliation{%
 Centre for Quantum Technologies, National University of Singapore, 3 Science Drive 2, S117543\\
}%
\author{Alexander Lohrmann}%
\affiliation{%
 Centre for Quantum Technologies, National University of Singapore, 3 Science Drive 2, S117543\\
}%
\author{Alexander Ling}
\affiliation{%
 Centre for Quantum Technologies, National University of Singapore, 3 Science Drive 2, S117543\\
}%
 \affiliation{Physics Department, National University of Singapore, 2 Science Drive 3, S117542}

\date{\today}

\begin{abstract}
We present an optical design where polarization-entangled photon pairs are generated within two $\beta$-Barium Borate crystals whose optical axes are parallel. This design increases the spatial mode overlap of the emitted photon pairs enhancing single mode collection without the need for additional spatial walk-off compensators. The observed photon pair rate is at least \SI[per-mode=symbol]{65000}{pairs/s/mW}  with a quantum state fidelity of $99.53\pm0.22$\% when pumped with an elliptical spatial profile.

\end{abstract}

\pacs{Valid PACS appear here}
\maketitle


Photon pair sources based on spontaneous parametric down-conversion (SPDC) are a standard tool in quantum optics when utilizing entanglement\cite{kiess93,bouwmeester1997experimental,jennewein00}. The SPDC process takes place in the presence of a nonlinear crystal where a pump photon ($p$) can, under conservation of energy and momentum, spontaneously split into two strongly correlated lower energy photons (signal, $s$; idler, $i$)\cite{coinc70}. Significant effort has been devoted towards developing bright entangled photon pair sources and a variety of source designs have been reported utilizing both critical\cite{kwiat95,kwiat1999ultrabright,kurtsiefer01,trojek2008collinear} and non-critical phase-matching techniques\cite{kim2006phase,steinlechner2013phase}. Increasingly such light sources are being deployed outside the laboratory \cite{marcikic2006free, ursin07,tang2016generation,yin2017satellite} where environmental considerations play a role in the selection of the phase-matching technique.

Non-critical phase-matching provides the brightest photon pair sources by providing access to a large nonlinearity \cite{steinlechner2012high} and longer interaction lengths due to the absence of spatial walk-off \cite{steinlechner2013phase,steinlechner2014efficient} while requiring very good temperature stability and often uses interferometric designs. When such requirements  (such as temperature stability) cannot be met, critical phase-matching (CPM), where the tilt angle of the crystal optical axis is used to fulfill the momentum conservation, can be applied.

A common feature in all source designs based on CPM-SPDC using bulk crystals is the spatial walk-off that any light polarized parallel to the crystal optical axis (extraordinary) will undergo. This effect becomes paramount when the source consists of more than one crystal for entangled light generation (typically, when using type-I CPM). This spatial displacement eventually restricts the final mode overlap of SPDC emission coming from multiple nonlinear crystals. The standard source geometry in this case is the so-called \textit{crossed-crystal} configuration comprising of two nonlinear crystals whose optical axes are rotated 90$\degree$ with respect to each other\cite{kwiat1999ultrabright,trojek2008collinear,rangarajan2009optimizing}.

  In type-I phase-matching ($e_p$ $\rightarrow$ $o_{s}o_{i}$; $e$: extraordinary, $o$: ordinary) this crossed-crystal configuration can be used to create the maximally entangled Bell states. Furthermore, it allows both collinear or non-collinear emission, while the wavelengths can be readily tuned. However, the fact that each crystal yields a pump walk-off in orthogonal directions (e.g., vertical in the first crystal and horizontal in the second crystal) results in a non-optimal spatial overlap of the two SPDC modes. Entangled light can only be collected from the region of optimum overlap, requiring additional birefringent crystals for spatial walk-off compensation. In addition to the spatial mismatch of the SPDC modes, the effective pump power is limited as the two crystals need to be pumped with orthogonal polarization components. We present a polarization entangled photon pair source using collinear type-I CPM uniaxial crystals whose optical axes are parallel. This \textit{parallel-crystal} configuration improves the spatial overlap of the target SPDC modes emerging from the two crystals. This relaxes the alignment requirements while improving the overall rate of detection for entangled photons pairs. Furthermore, the source design uses fewer birefringent elements reducing the physical footprint.
\begin{figure*}[t!]
\centering
\includegraphics[scale=0.85]{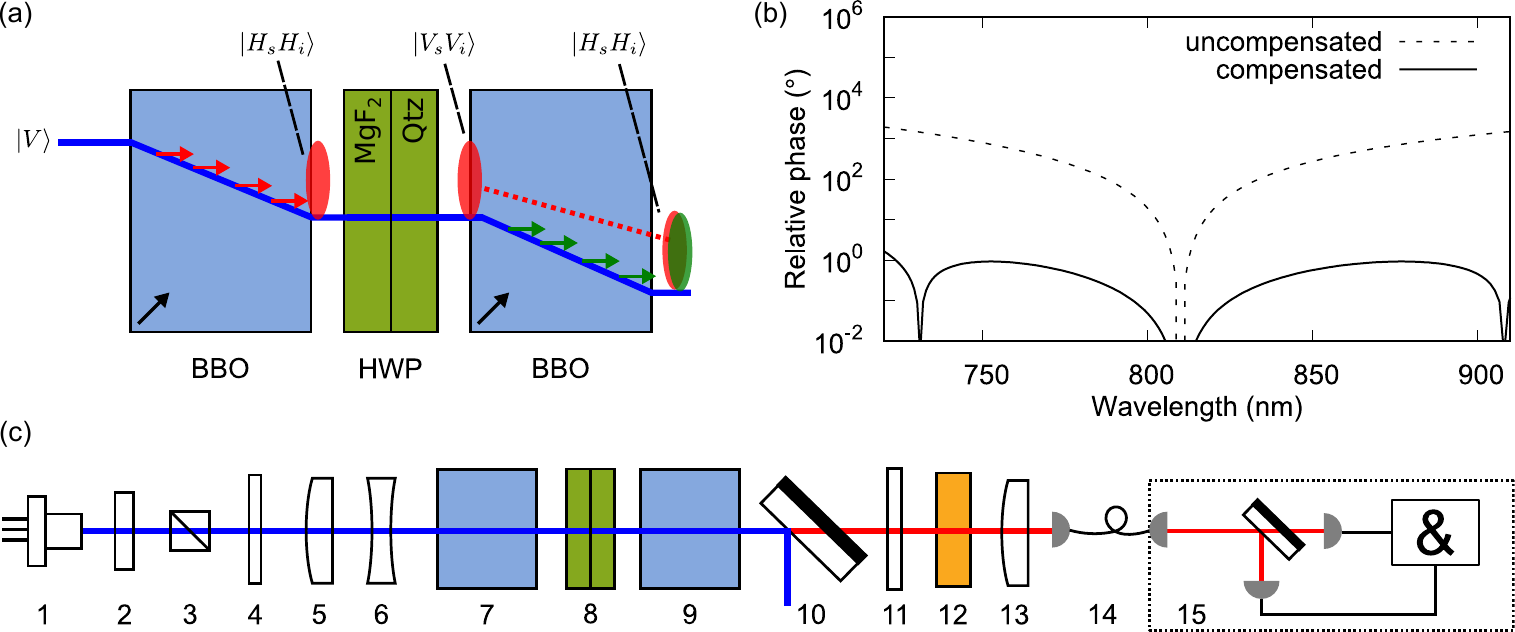}

\caption{Schematic for the source design. (a) Spatial self-compensation of the emitted photon pairs. The BBO optical axes are both pointing upwards as indicated by the black arrow. SPDC light is created along the pump path in both crystals and the red (green) ellipse indicates the shape of the SPDC emission generated in the first (second) BBO crystal. An achromatic half-wave plate (with magnesium fluoride (MgF$_2$) and quartz (Qtz) components) rotates the polarization of the photons born in the first BBO. (b) Theoretically calculated relative phase between $\ket{V_sV_i}$ and $\ket{H_sH_i}$ with (solid line) and without (dashed line) the phase compensation crystal. For BBO length of \SI{5}{\milli\metre}, a YVO$_4$ length of \SI{3.12}{\milli\metre} can act as the compensator. (c) Overview over the experimental setup, 1: laser diode, 2: fluorescence filter, 3: Glan-Taylor polarizer, 4: half-wave plate, 5: convex lens, 6: concave lens, 7: BBO, 8: achromatic half-wave plate, 9: BBO, 10: dichroic mirror, 11: long-pass filter, 12: phase compensation crystal, 13: position of the analyzer when performing the single polarizer measurement (see Fig.~\ref{fig:set2}(b)), 14: collection lens, 15: single-mode fiber, 16: photon separation via a dichroic mirror followed by coincidence detection. (Color online).}
\label{fig:set1}
\end{figure*}

The source design presented in this work avoids the characteristic drawbacks of the crossed-crystal source (lower effective pump power and two-dimensional walk-off) by using two crystals with parallel-aligned optical axes, as shown in Fig.~\ref{fig:set1}(a). This allows both crystals to utilize the full pump power in the SPDC process. As the photon pairs in both crystals are generated in the same polarization state ($\ket{H_sH_i}$), the SPDC pairs from the first crystal need a 90$\degree$ polarization rotation\cite{zhang2015experimental}. The final quantum state becomes,

\begin{equation}
\ket{\Phi^{\pm}} = \frac{1}{\sqrt{2}} \left( \ket{H_sH_i} \pm e^{i\Delta\varphi}\ket{V_sV_i} \right),
\end{equation}
where $\Delta\varphi$ denotes the phase difference between photon pairs born in the first and second crystals. To realize the rotation an achromatic half-wave plate is inserted in between the two crystals. This half-wave plate is designed to have no effect on the pump polarization. When the pump light travels through both crystals it undergoes walk-off in the vertical direction. Similarly the SPDC photons from the first crystal walk off in the same direction when traversing the second crystal.
\begin{figure}[b]
	\begin{center}
	\includegraphics[scale=0.85]{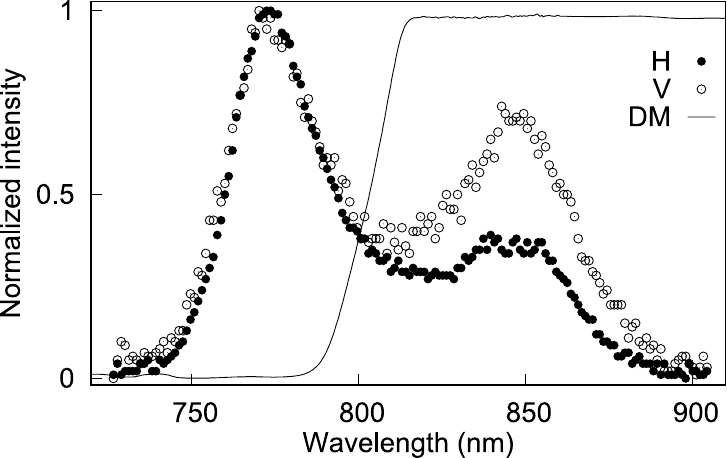}
	\caption{Spectra for horizontally (black) and vertically (white) polarized photons. Solid line: dichroic mirror spectral response. More information about the dichroic mirror response can be found in \cite{septriani2016thick}. Resolution of the spectrometer: \SI{2}{\nano\metre}. The intensity mismatch of the idler peaks for the two different polarization states is due to the polarization dependence of the grating spectrometer used.}
	\label{fig:spectrum}
    \end{center}
\end{figure}
As depicted in Fig.~\ref{fig:set1}(a) this crystal configuration reduces the necessity of any additional spatial compensation crystals since the photon pairs emitted from both crystals overlap significantly in the same plane when exiting the second crystal (red and green ellipses). Note that the walk-off is a function of the wavelength and the crystal angle cut. In our experimental design using a pump wavelength of \SI{405}{\nano\metre} and $\beta$-Barium Borate (BBO) crystals with an angle cut of \SI{28.8}{\degree}, the walk-off introduces a 6\% mismatch of the photon pair emission along the vertical direction, independent of crystal length. If necessary, this mismatch can be compensated by using a single birefringent crystal, but is deemed sufficiently small in this work and neglected. Due to the spatial self-compensation and the maximal use of the pump power in both crystals, the parallel crystal configuration promises a significant increase in brightness over the traditional crossed-crystal configuration.

In order to observe a Bell state, the relative phase $\Delta \varphi$ between photon pairs generated in the first and second BBO crystals should have a negligible wavelength dependence\cite{trojek2008collinear, kim2000high}. 
Photon pairs generated in the first (second) crystal will accumulate a phase $\varphi_1$ ($\varphi_2$) as they travel through the dispersive medium. The contribution to $\varphi_1$ arises from the photon pairs traveling through the half-wave plate and the second BBO crystal. Additionally, it should be noted that these photon pairs have extraordinary polarization after the half-wave plate (HWP). The contribution to $\varphi_2$ is similar to $\varphi_1$, but differs where the phase through the first BBO crystal and the half-wave plate is picked up by the parent photon.
\begin{figure*}[t!]
	\begin{center}
	\includegraphics[scale=0.85]{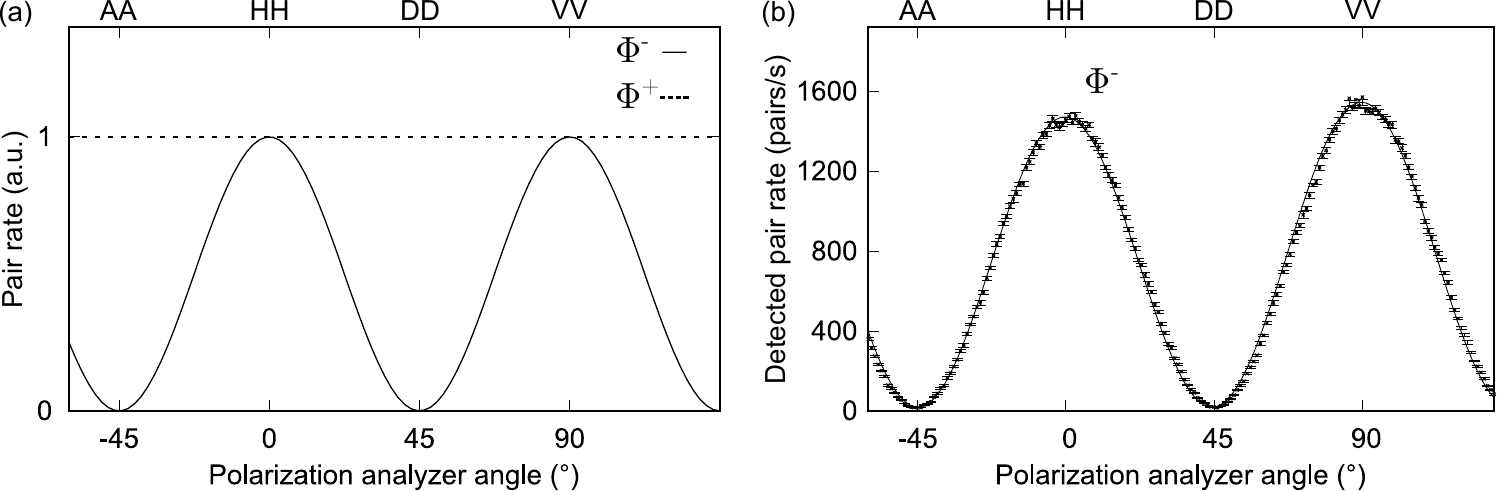}
	\caption{(a) Theoretical pair rates for both $\Phi^+$ and $\Phi^-$ states when utilizing a single polarizer analyzer. This measurement technique can distinguish between the maximally entangled states $\Phi^+$ (zero contrast curve) and $\Phi^-$ (full contrast curve). (b) Detected photon pair curve using a single polarizer (installed after the temporal compensator, see item 13 in Fig.~\ref{fig:set1}(c)) measurement at a pump power of \SI{0.1}{\milli\watt}. The peak at 0 (90) degree corresponds to horizontal (vertical) polarization. The minimum at -45 (45) degrees corresponds to anti-diagonal (diagonal) polarization. The fidelity calculated from the fit was $99.53\pm0.22\%$.}
	\label{fig:set2}
    \end{center}
\end{figure*}

Without any additional birefringent elements, the relative phase will show a strong wavelength dependence. This wavelength dependence can be suppressed by the addition of an a-cut single yttrium orthovanadate crystal (YVO$_4$) with the correct orientation and length (\SI{3.12}{\milli\metre} in our case). Therefore $\Delta \varphi $ can be calculated with the following equation (assuming the Sellmeier equations for BBO\cite{eimerl1987optical}, YVO$_4$\cite{shi2001measurement}, and MgF$_2$\cite{dodge1984refractive} and Quartz\cite{ghosh1999dispersion} for the half-wave plate),
\begin{gather}
\Delta \varphi = \varphi_1 - \varphi_2 ,\\ \varphi_1=  \varphi^ \text{HWP}_{s} +\varphi^ \text{HWP}_{i} +\varphi^{\text{BBO}}_{s,e}+ \varphi^{\text{BBO}}_{i,e}+\varphi^{ \text{YVO}_4}_{s,o}+\varphi^{ \text{YVO}_4}_{i,o},\\ \varphi_2=\varphi^{\text{BBO}}_{p,e} + \varphi^{\text{HWP}}_p + \varphi^{ \text{YVO}_4}_{s,e}+ \varphi^{ \text{YVO}_4}_{i,e}.
\end{gather}

Due to the narrow bandwidth of the pump laser used in this work ($\leq$ \SI{160}{\mega\hertz}) all terms containing pump photons can be treated as a constant.

The relative phase is first evaluated without the phase compensation crystal to show strong wavelength-dependence (Fig.~\ref{fig:set1}(b), dashed line). For clarity, the constant offset at the degenerate wavelength (\SI{810}{\nano\metre}) was subtracted. After adding the phase compensation crystal, the relative phase shows a negligible wavelength dependence (Fig.~\ref{fig:set1}(b), solid line). With this amount of compensation, the maximally entangled state $\ket{\Phi^-} = \left( \ket{H_s H_i} - \ket{V_s V_i}\right)/\sqrt{2} $ can be achieved.

\begin{figure*}[tbp]
	\begin{center}
	\includegraphics[scale=0.85]{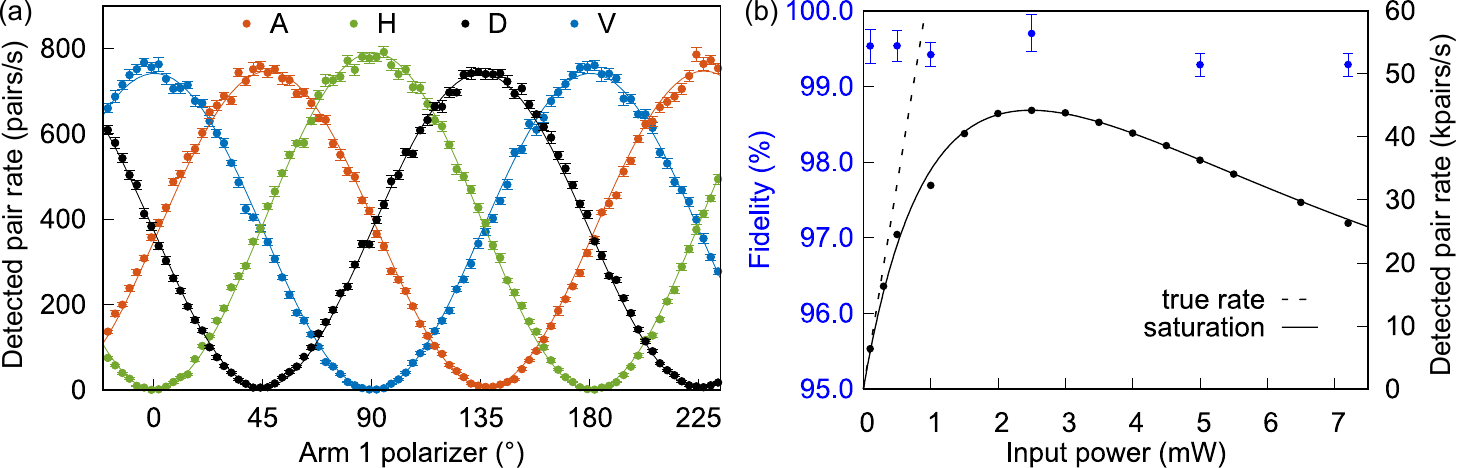}
	\caption{(a) Photon pair coincidences obtained using two polarizers. For this measurement, the dichroic mirror was replaced with a 50:50 beam splitter (to avoid wavelength-dependent phase-shifts). Polarization analyzers were placed in both arms after the beam splitter. Single-mode fiber coupling was performed after the polarization analyzer. While fixing the polarization analyzer setting in one arm ($A, H, D, V$), the polarizer in the other arm was swept and the coincidence events were recorded. (b) Power dependence of the fidelity (blue, left axis) and the detected pair rate (black, right axis) with the polarization analyzer in place. The source saturates the passively-quenched detectors at \SI{2}{\milli\watt} of pump power. The error bars for the experimental rate data points are smaller than the symbols. The solid black line is a guide to the eye indicating the saturation of the coincidence rate. The rate assuming linear power scaling is depicted by the dashed black line. (Color online).}
	\label{fig:set3}
    \end{center}
\end{figure*}

The experimental setup is shown in Fig.~\ref{fig:set1}(c). Collimated light from a grating stabilized laser diode with a central wavelength of \SI{405}{\nano\metre} was used to pump the nonlinear medium. Undesirable pump spectral components were removed by a fluorescence filter. The pump polarization state was prepared by a Glan-Taylor polarizer and a half-wave plate. In principle, these polarizing elements are not necessary as the laser output is highly polarized.

The pump beam has an elliptical spatial profile, with the major axis aligned in the walk-off direction. The ratio of the major and minor axes was 2:1. The pump focus (along the minor axis) was set in between the two \SI{5}{\milli\metre} thick crystals. The full-width half-maxima of the pump beam at the focus were \SI{133}:\SI{63}{\micro\metre} while the target signal and idler wavelengths were centred on \SI{776}{\nm} and \SI{847}{\nm}, respectively. A tailored achromatic half-wave plate rotates the polarization of the SPDC photons generated in the first crystal, while the pump polarization remains unaffected. Dichroic mirrors remove the excess pump.

The photon pairs were collected into a single-mode fiber (SMF) using an achromatic collection lens. The collection mode full-width half-maximum was set at \SI{53}{\micro\metre} (in between the two BBO crystals) to enhance coupling efficiency. The collected photon pairs were split with respect to their wavelengths using a dichroic mirror, and guided into passively-quenched avalanche photodiodes (APDs) capable of single-photon detection. The single photon detection events from two APDs were checked for timing correlations using a coincidence time window set at \SI{4}{\nano\second}.

The entangled photon pair rate was measured without additional interference or polarization filters and a pump power of \SI{0.1}{\milli\watt} was used to remain within the linear detection regime of the single-photon detectors. The normalized detected pair rate was \SI[per-mode=symbol]{65000}{pairs/s/mW}. In comparison to the brightest reported CPM crossed-crystal source (\SI[per-mode=symbol]{27000}{pairs/s/mW})\cite{trojek2008collinear} using similar detector efficiencies ($\approx$55~\%), this corresponds to a 2.4 times increase in brightness. This improvement was achieved despite using crystals that were only one-third in length.

The heralding efficiency (coincidences-to-singles ratio) without correcting for detector efficiency and transmission losses after subtracting for dark-counts, was $27.0\pm1.0\%$ and $22.0\pm1.0\%$ for signal and idler photons, respectively. Note that these efficiencies are in the same range as when the pump is spatially filtered by a SMF. Spectral information of the SPDC photons is shown in Fig.~\ref{fig:spectrum}.

We test the correlations using a single polarization analyzer placed after the phase compensation crystal (see the position of the polarization analyzer in Fig.~\ref{fig:set1}(c)). The position of this analyzer allows it to act on both signal and idler photons at the same time. The use of a single polarization element to perform partial tomography is not common in the literature; it is appropriate in this work because the photons in the prepared state $\Phi^-$ are always co-polarized in one basis ($H/V$), while anti-polarized in the other linear basis ($D/A$). This leads to a unique signature between the states $\Phi^-$ and $\Phi^+$. For $\Phi^-$, the curve will show high contrast between maxima and minima, while $\Phi^+$ will show a constant pair rate for any polarization setting.

In analyzing the generated two-photon state, we make the following assumptions. First, we neglect the contribution of the $\ket{HV}$ and $\ket{VH}$ components to the total state. This is due to the phase-matching conditions preventing orthogonally polarized $s$ and $i$ photons, and the high precision of the half-wave plate rotating the down-converted photons from the first crystal. Then the state can be simplified to a two-level system comprising $\ket{HH}$ and $\ket{VV}$ states. The description for our generated quantum state becomes,
\begin{multline}
\rho = p\bigg(\sqrt{x}\ket{HH} + e^{i\theta}\sqrt{1-x}\ket{VV}\bigg) \\ \bigg(\sqrt{x}\bra{HH} + e^{-i\theta}\sqrt{1-x}\bra{VV}\bigg) \\ + \frac{(1-p)}{2}\bigg(\ket{HH}\bra{HH} + \ket{VV}\bra{VV}\bigg),
\end{multline}
with $p$, $x$ and $\theta \in \mathbb{R}$. The conditions $0 < p$, $x < 1$ ensure the state is normalized. $p$ denotes the purity, $x$ describes the balance of the Bell state and $\theta$ is the relative phase. The projection of this state to the single polarizer measurement is directly fitted to the normalized visibility curve of the measurement, and the parameters $p$, $x$ and $\theta$ can be extracted. By bootstrapping the fit procedure we obtain uncertainty intervals for the fitting parameters. With the parameters extracted from the fit, we reconstruct $\rho$ and calculate the Bell state fidelity, $F \left(\rho, \sigma\right) =  \text{Tr}\left( \sqrt{\sqrt{\rho} \sigma \sqrt{\rho}} \right)$, where $\sigma = \ket{\Phi^-}\bra{\Phi^-}$.

The correlation curve was measured and is presented in Fig.~\ref{fig:set2}(b). From this correlation curve a Bell state fidelity of $99.53\pm0.22\%$ was extracted. For comparison, the standard measurement using two polarization analyzers, each independently acting on the signal and idler photons, was also performed (see Fig.~\ref{fig:set3}(a)). We measured the correlations in the horizontal/vertical (${H/V}$), diagonal/anti-diagonal (${D/A}$) linear bases and the left/right (${L/R}$) circular basis. In this case, we obtained visibilities of $V_{H/V} = 99.70\pm0.11$, $V_{D/A} = 98.32\pm0.18$ and $V_{L/R} = 98.6\pm0.3$\%. From these visibility values, we can deduce the fidelity using the equation $F= \sqrt{\frac{1}{4}(1+V_{H/V}+V_{D/A}+V_{L/R})}$\cite{wagenknecht2010experimental}. We calculate a fidelity of $99.6\pm0.4\%$ which is in excellent agreement with the single polarizer measurement.

Experiments that utilize such a high brightness source would not be limited by pump power but by detector performance. We show in Fig.~\ref{fig:set3}(b) that the quantum state fidelity remained above 99$\%$ even when the passively-quenched detectors were heavily saturated. It is noteworthy that high brightness, good collection efficiency and near-unit quantum state fidelity were reported when the source was pumped directly from a laser diode without any spatial filtering. This is in agreement with early theoretical work on elliptical pump beams for parametric gain \cite{kuizenga1972optimum} and general statements for other nonlinear optical parametric processes \cite{freegarde1997general}. Work is ongoing to identify the optimal crystal lengths and beam parameters for achieving higher brightness.

Another direction is to investigate whether brightness and entanglement quality are conserved when the emitted photon pairs are collected through a field stop instead of a single-mode fiber. The combination of field stop collection with direct pumping from a laser diode without spatial filtering could lead to a new class of bright entangled light sources useful in both fundamental science and downstream applications.

\section*{Funding}
National Research Foundation (NRF); Competitive Research Programme (CRP); Award No. NRF-CRP12-2013-02; Ministry of Education (MOE); Research Centres of Excellence (RCE).

\section*{Acknowledgments}
The use of a single polarizing element for partial tomography was first suggested by W. Morong. The authors thank A. Stolk for assistance in the single polarizer method derivation.

\bibliography{bibliography}

\end{document}